\documentclass[a4paper, 10pt, final]{article}
\usepackage[latin1]{inputenc}       
\usepackage[T1]{fontenc}            
\usepackage[francais]{babel}         
\usepackage{url}                    
\usepackage{amsmath}
\usepackage{times}
\usepackage{a4wide}
\usepackage{amssymb}
\usepackage{tabularx}
\usepackage[dvips]{graphics}
\usepackage{fancybox}
\usepackage{epsfig}
\usepackage{graphicx}
\usepackage{color}
\usepackage{floatflt,amssymb}
\usepackage{moreverb} 
\usepackage{multirow} 
\usepackage[all]{xy}  
\usepackage{textcomp} 
\usepackage{hyperref} 
\usepackage[right]{eurosym}

\begin{document}

\title{Compte-Rendu\\Conférence ECMDA 2007\\ \normalsize{11-14 juin, Haifa, Israël}}
\date{\today}

\author{
       Beno\^{\i}t Combemale\\
       \textit{IRIT CNRS (UMR 5505)}\\
       2, rue Charles Camichel\\
       F-31071 Toulouse Cedex 7\\
       \normalsize{\url{http://combemale.perso.enseeiht.fr}}
}

\maketitle

\abstract{Ce document reprend mes notes personnelles, sur un style "informel", prises au cours de la conférence ECMDA 2007. Il mentionne l'intégralité du programme de la conférence (titre de l'article et nom de la personne qui a présenté) en détaillant certaines des présentations.}

\tableofcontents

\newpage

\section{Introduction}

La conférence européenne ECMDA-FA'07 (\url{http://www.haifa.il.ibm.com/conferences/ecmda2007/}) sur l'Ingénierie Dirigée par les Modèles a eu lieu le 13 et 14 juin 2007 à Haifa (Israël).
Cette conférence est découpée en 2 types de session : Applications (plutôt industrielles) et Foundations (plutôt académiques). Les actes sont édités par Springer dans le LNCS Vol. 4530.
Le programme de la conférence est accessible à l'adresse \url{http://www.haifa.il.ibm.com/conferences/ecmda2007/program.html} et le résumé de chacune des présentations à l'adresse \url{http://www.haifa.il.ibm.com/conferences/ecmda2007/abstracts.html}.

Adossé à cette conférence ont eu lieu le 11 juin un workshop informel sur la rencontre entre académiques et industriels et le 12 juin un workshop sur la tracabilité des modèles (\url{http://modelbased.net/ecmda-traceability/}) dont les actes sont disponibles à l'adresse \url{http://modelbased.net/ecmda-traceability/papers07/ECMDA-TW-07-Proceedings-Final.pdf}.

\section{ECMDA : Main Conference (June 13, 2007)}

\subsection{Keynote : Stuart Kent (Microsoft)}

\subsubsection*{Pragmatics of Model Driven development}

Après avoir présenté les notions de DSL et de DSD (Domain-Specific Development) tels qu'ils sont définis chez Microsoft, Stuart Kent a présenté le pragmatique de la plateforme Visual Studio pour le développement spécifique à un domaine. 
A travers un exemple d'application (réservation de chambres d'hôtel), il a présenté la plate-forme selon les points suivants :

\begin{itemize}
\item Validation de contraintes : utilisation du langage LINQ, un langage proche d'OCL avec des primitives pour la génération de messages d'erreurs.
\item Modèles complexes : possibilités de combiner différents diagrammes (de même type ou pas) correspondants à différents points de vue ou à différents niveaux de hiérarchisation. 
La plate-forme supporte le référencement de fragments de modèle à partir d'opérateurs comme diff., merge\ldots.
\item Mappings (génération de code, reverse, model to model) : le but est de pouvoir travailler sur des parties générées. On définit pour cela sur le code ou le modèle les parties qui sont maîtres (lecture/écriture) et celles qui sont esclaves (lecture uniquement). Une limite toutefois : le problème du round-tripping.
\item Personnalisation : la personnalisation peut se faire sur les générateurs d'outils, sur les générateurs de code ou sur le code généré. Le versionning devient alors compliqué : Tool platform vendor (v1, v2...), tool vendor (v1.1, v1.2...v2.1....), tool customizers (v1.1.1, 1.1.2....) et application builder (v1.1.1.1....). Cela pose alors le problème de qui peut personnaliser quoi ? La plate-forme doit permettre de le définir clairement.
\item Intégration du processus : la plate-forme essaye de combiner l'approche DSL avec les méthodes agiles.
\end{itemize}

\subsection{Session A : Applications}

\subsubsection*{Model Transformation from OWL-S to BPEL Via SiTra (Gareth Howells, Kent  University)}

L'exposé a porté principalement sur la présentation du langage de transformation SiTra (\textit{Simple Transformer}); l'application OWL2BPEL détaillée dans l'article n'étant que brièvement évoquée.

Les hypothèses de conception de ce langage de transformation sont :
\begin{itemize}
\item les utilisateurs ne veulent pas utiliser une nouvelle approche autre que celle des langages de programmation,
\item ils ne veulent pas appréhender de nouveaux concepts de programmation.
\end{itemize}

Ainsi, SiTra a pour objectif de permettre d'écrire une transformation par un langage de programmation (java\ldots) et fournit pour cela des librairies : la classe \emph{SimpleTransformer}, les interfaces \emph{Rule} et \emph{Transformer}, et un certain nombre de méthodes (\emph{check} pour vérifier si la transformation peut être appliquée, \emph{build} pour construire un élément du modèle cible, \emph{setProperties} pour assigner des valeurs à des propriétés du modèle cible\ldots) à spécialiser pour chaque transformation.

L'exécution de la transformation revient alors à créer une instance de \emph{Transformer}, à enregistrer la transformation, puis à appeler chacune des règles quand la méthode \emph{check} renvoie \emph{true} (ie. l'objet n'a pas encore été traité).

Les fonctionnalités principales de la version actuelle sont la définition de règles simples, la réutilisation de règles (composition ET/OU), l'héritage de règles et l'invocation implicite/explicite de règles.

Les limitations actuelles de la plate-forme sont  l'invocation récursive d'une règle et le choix de la règle à appliquer pour un même objet.

\subsubsection*{Towards a Model-Driven Approach to automatic BPEL (Xiaofeng Yu, Nanjing University par télé-conférence)}

Présentation d'un mapping en QVT entre EDOC-CCA (profil UML de l'OMG pour EDOC qui permet de modéliser la structure et le comportement de composants et leurs orchestrations) et BPEL (langage de workflows exécutable). Le but est d'utiliser EDOC-CCA pour spécifier de manière abstraite l'orchestration puis générer automatiquement le workflow BPEL.

\subsubsection*{A Model-Driven Software Factory using DSL (présenté par Anneke Kleppe, Twente University)}

L'exposé a commencé par une introduction au framework Microsoft :
Les développeurs de DSL définissent un \emph{domain model} à partir duquel ont peut générer un composant qui s'installe dans l'IDE pour le développement d'applications.
Les développeurs d'applications définissent des \emph{application model}, conformes au modèle de domaine et génère ensuite semi-automatiquement le code de leur application.

Le reste de la présentation était un retour d'expérience sur l'utilisation de ce framework pour le développement de l'usine logicielle SMART-Microsoft. Les conclusions principales sont :
\begin{itemize}
\item d'utiliser des petits DSL et accepter qu'ils ne prennent pas tout en compte,
\item un modèle ne doit pas tenir compte d'une technologie particulière. C'est la transformation Model2Code qui doit introduire cette technologie.
\end{itemize}

\subsection{Session B : Foundations}

\subsubsection*{A Practical Approach to Model Extension (Jean Bézivin, INRIA)}

L'exposé a commencé par une présentation de la plate-forme AMMA (ATM, AMW, AM3\ldots) puis a montré la nécessité d'avoir des mécanismes d'extension de modèle (+ notion de treillis de modèles).
L'exemple pris pour cela est le méta-modèle des réseaux de Petri que l'on souhaite étendre pour ajouter le marquage courant au cours d'une exécution.

Les auteurs proposent une formalisation de la relation "extensionOf" (un sous ensemble clair de l'opérateur "merge" introduit par l'OMG) et propose trois moyens de l'implémenter : concaténation de descriptions textuelles en KM3, un modèle de lien (weaving) entre les 2 parties du méta-modèle ou une approche par transformation de modèle qui complète le premier par les informations du second.

\subsubsection*{Towards the Generation of a Text Based IDE from a language Metamodel (Anneke Kleppe, Twente University)}

Après avoir introduit les principales différences entre un méta-modèle et une grammaire (références conteneurs, mots-clés absent du metamodèle\ldots), l'auteur propose une approche pour définir des outils textuels (basé sur une BNF) à partir d'un méta-modèle.

Elle définit la notion de \emph{Parse Model} (PM) qui rajoute au modèle de la syntaxe abtraite (ASM), à travers une transformation ASM2PM, les mots-clés permettant le parsing de texte (Rq: il peut y avoir plusieurs PM pour un ASM). A partir de là, une transformation PM2BNF fait le pont vers les outils génératifs de l'ingénierie des langages.
Les principales conclusions sont :
\begin{itemize}
\item des outils textuels peuvent être générés à partir d'un méta-modèle,
\item il faut une séparation claire entre la syntaxe abstraite et une ou plusieurs syntaxes concrètes.
\end{itemize}

\subsubsection*{Human Comprehensible and Machine Processable Specifications of Operational Semantics (Markus Scheidgen, Berlin University)}

Les auteurs présentent dans cet article une méthode pour décrire la sémantique opérationnelle d'un méta-modèle. 
Ils proposent de définir pour une classe statique (instance de la métaclasse contenant les informations statiques), $n$ classes d'exécution, toutes instances de la métaclasse héritant de la première et rajoutant les informations dynamiques, i.e. évoluant au cours de l'exécution.
Ils proposent ensuite de définir des diagrammes d'activités pour définir les transitions possibles d'un état à l'autre d'un modèle. L'opérationalisation est exprimée a l'aide d'actions de base de création, modification, suppression\ldots

Associé à cette approche, ils ont définis un interpréteur générique permettant l'exécution de modèles définis à partir d'un méta-modèle complété d'une telle description de la sémantique.

\subsection{Session C : Applications}

\subsubsection*{Improving the Interoperability of Automative Tools by Raising the Abstraction from Legacy XMI Formats to Standardized Meta-Models (Mark Brokens, Carmeq GmbH)}

\subsubsection*{Model Re-engineering from Traces to Validate Distributed Systems - An Industrial Case Study (Andreas Ulrich, Siemens AG)}

\subsubsection*{Adopting Model-Driven Development in a Large Financial Organization (Yael Shaham-Gafni, Metaphor Vision Ltd.)}

\section{ECMDA : Main Conference (June 14, 2007)}

\subsection{Keynote : Andy Schürr (Darmstadt University of Technology)}

\subsubsection*{Model-Driven Development With Matlab Simulink \& Stateflow}

L'exposé a montré l'utilisation de l'atelier Matlab Simulink \& Stateflow (MSS) dans les projets Mate et Toolnet.

Le projet MATE (\url{http://www.moflon.org/success\_stories/mate.html}) avait pour objectif de définir des guidelines sous MSS et les vérifier sur les modèles définis. Le principe adopté est de passer par le MOF et la définition de SDM (diagramme d'activité UML + règles de ré-écriture de graphes) à l'aide de transformations définis avec l'environnement MOFLON (\url{http://www.moflon.org/}). Les apports par rapport à OCL sont d'avoir des propriétés plus simples à lire, de pouvoir les définir de manière graphique et avec un plus fort pouvoir d'expression (fermeture transitive ...).

Le projet TOOLNET (\url{http://www.moflon.org/success\_stories/toolnet.html}) permet d'établir des liens de traçabilité entre des exigences et les modèles MSS. Ils définissent pour cela des spécifications bi-directionnelles (en utilisant le langage de transformation de graphe TGG, sous ensemble de QVT) et génèrent tous les différents scénarios.

L'exposé se termine par une description de l'atelier MOFLON (MOF, DiaGen, OCL, SDM, TGG, Velocity) et une comparaison par rapports aux autres (AMMA, GME, Fujaba\ldots).

Les problèmes ouverts :
\begin{itemize}
\item certains concepts UML (associations qualifiées\ldots),
\item intégrer la définition de profil UML,
\item contraintes OCL : vérification de contraintes incrémentales, fermeture transitive, pattern-matching\ldots
\item définition formelle des transformations de modèles locales avec SDM,
\item fusionner TGG avec QVT relationnel,
\item l'intégration avec le framework de génération d'éditeur Diameta (\url{http://www.moflon.org/success\_stories/diameta.html}).
\end{itemize}

\subsection{Session D : ModelWare}

\subsubsection*{Is MDA Ready for Real Business? - A User Perspectives! (Bjorn Nordmoen, WesternGeco)}

Conclusions en vrac :
\begin{itemize}
\item on peut pas évaluer l'amélioration de la productivité,
\item la création de profils est trop compliquée.
\end{itemize}

\subsubsection*{Adopting MDA in a Small Enterprise (Regis Vogel, IHG)}

\subsubsection*{Experimental Validation of the Effects of Model Driven Development on the Software Lifecycle (Yael Dubinsky, IBM Haifa Research Lab)}

\subsubsection*{Modelware Discussion}

\subsection{Session E : Foundations}

\subsubsection*{Scenarios of Traceability in Model to Text Transformations (Goran Olsen, SINTEF)}

Présentation de l'outil MOFscript implémentant le langage de transformation de même nom. Il a été développé sous la forme d'un plugin Eclipse, dans le cadre de Modelware.

La présentation souligne la nécessité d'avoir des liens de traçabilité dans le cadre de transformations modèle à texte (traçabilité end-to-end, spécifications des transformations différentes du développement de système par les ingénieurs, modifications manuelles du code généré\ldots).
Dans MOFscript :
\begin{itemize}
\item la transformation a des références vers les éléments du modèle,
\item chaque référence est utilisée pour générer un fichier texte.
\end{itemize}
De cette manière, le prototype permet d'avoir une vision de la trace pour chaque élément du modèle, une couverture des éléments utilisés pour générer\ldots

\subsubsection*{An Algebraic view on the Semantics of Model Composition}

Formalisation de la composition de modèles à différents niveaux (syntaxique, sémantique, méthodique) et études de différentes propriétés à chacun des niveaux.

\subsubsection*{Execution of Aspect oriented UML Models (Pablo Sanchez, University of Malaga)}

Hypothèse de départ : les modèles exécutables sont une réalité mais pas encore de module de conception d'applications basé sur le "cross-cutting" :
\begin{itemize}
\item[$\Rightarrow$] expression du comportement par des diagrammes d'activités,
\item[$\Rightarrow$] transformation en Xpath et ATL pour gérer le cross cutting.
\end{itemize}

\subsection{Session F : Foundations}

\subsubsection*{Templateable Metamodel for Semantic Variation Points (Arnaud Cucurru, CEA LIST)}

Les auteurs s'intéressent aux points de variation sémantique (PVS) dans les métamodèles.  
Ces PVS consistuent une
information essentielle dans le domaine de l'ingénierie des langages mais sont
aussi présents dans l'IDM où ils ne sont pas clairement identifiés et sont
décrits le plus souvent informellement dans la documentation associée aux
métamodèles.  La spécification d'UML 2 est un bon exemple de
cet état de fait.  Ainsi, le MOF ne propose aucun mécanisme pour identifier ni
traiter les PVS.

Les auteurs proposent de s'inspirer du mécanisme de généricité présent dans
les langages de programmation pour paramétrer un métamodèle, aussi bien au
niveau des paquetages que des méta-classes. 
La notation proposée s'appuie sur les templates d'UML 2 et
est illustrée sur les machines à états d'UML 2 en
s'intéressant à deux points de variation sémantique : la sélection des
événements et la sélection des transitions.  Chacun de ces deux PVS est
décrit par une méta-classe, est utilisé comme paramètre de généricité du
paquetage des machines à état et de la méta-classe \emph{State} concernée par
ces PVS.  Deux réalisations concrètes de ces PVS sont proposées correspondant
à des politiques respectivement aléatoire et stochastique.

\subsubsection*{Constraints Modeling for (Profiled) UML Models (François Lagarde, CEA LIST)}

L'exposé part du constat qu'un stéréotype ajoute une information sémantique à un élément du modèle. L'augmentation du nombre de profils définis amène à ce poser des questions sur leurs comptabilités entre eux, i.e. définition d'un cadre sain pour appliquer plusieurs profils simultanément. 
Dans ce contexte, les auteurs proposent des mécanismes pour exprimer des contraintes portant sur l'application. Ces mécanismes ont été définis par des profils.

Quelques informations ont été données sur le prototype (plugin Eclipse) qui est en train d'être développé et les algorithmes utilisés pour évaluer les règles sur les modèles.

\subsubsection*{Efficient Reasoning About Finite Satisfiability of UML Class Diagrams with Constrained Generalization Sets (Azzam Maraee, Ben-Gurion University of Israël)}

\section{Conclusion}

La prochaine édition de cette conférence ce tiendra du 9 au 12 juin 2008 à Berlin. La date limite de soumission des articles est d'ores et déjà fixée au 5 février 2008.

\end{document}